# Resolution enhancement in digital holography by self-extrapolation of holograms

**Tatiana Latychevskaia[*] and Hans-Werner Fink**

*Physics Institute, University of Zurich, Winterthurerstrasse 190, CH-8057, Switzerland*

[*]tatiana@physik.uzh.ch

## Abstract

It is generally believed that the resolution in digital holography is limited by the size of the captured holographic record. Here, we present a method to circumvent this limit by self-extrapolating experimental holograms beyond the area that is actually captured. This is done by first padding the surroundings of the hologram and then conducting an iterative reconstruction procedure. The wavefront beyond the experimentally detected area is thus retrieved and the hologram reconstruction shows enhanced resolution. To demonstrate the power of this concept, we apply it to simulated as well as experimental holograms.

## 1. Introduction

Intrinsic to the principle of holography is that even a fraction of a hologram permits the reconstruction of the entire object [1-2]. However, in digital holography the overall achievable resolution in the reconstruction is believed to be limited by the size of the captured digital record [1-4]:

$$R = \frac{\lambda z}{\mathrm{N}\Delta_{\mathrm{S}}}, \qquad (1)$$

where $\lambda$ is the wavelength, z is the distance between object and screen, N is the number of pixels, and $\Delta_S$ is the pixel size of the detector. $N\Delta_S$ is thus the size of the hologram and hence the limiting aperture of the optical system. Therefore, in the case of digitally recorded holograms, the detector size appears to limit the achievable resolution.

We make use of the basic notion that the interference distribution recorded in a hologram is created by waves that are continuous and must persist beyond the experimentally recorded area. Consequently, it must be possible to self-extrapolate a truncated hologram beyond the intensity distribution actually recorded in an experiment. We will show how self-extrapolation of holograms recorded in an inline scheme [1-2, 5-6] can be achieved by applying an iterative routine. After several iterations, the outer part of the hologram, which was initially not available from the experimental record, is retrieved. The availability of these emerging higher-order fringes in such self-extrapolated holograms leads to an improved resolution in the object reconstruction.

## 2. Method

The method of self-extrapolation of a hologram requires a fraction of a hologram of size $N_0\Delta_S \times N_0\Delta_S$, where $N_0$ is the number of pixels, referred to below as $H_0$. The self-extrapolated hologram shall have a final extended size of $N\Delta_S \times N\Delta_S$ where the number of pixels $N>N_0$ and the pixel size of the detector $\Delta_S$ remains unchanged. Self-extrapolation is achieved by applying an iterative procedure [7-10] which basically boils down to the field being propagated back and forth between screen- and object plane, whereby constraints, as illustrated in Fig. 1, are applied. The reconstructed transmission function in the object plane is given by the solution of the back propagation integral [9, 11]

$$t\left(\vec{r}_O\right)=\frac{i}{\lambda}r_O\exp\left(-ikr_O\right)\iint U\left(\vec{r}_S\right)\frac{\exp(-ik\left|\vec{r}_S-\vec{r}_O\right|)}{\left|\vec{r}_S-\vec{r}_O\right|}\mathrm{d}\sigma_S, \qquad (2)$$

where $U\left(\vec{r}_S\right)$ is the complex-valued field distribution in the screen (hologram) plane, $t\left(\vec{r}_O\right)=1+o\left(\vec{r}_O\right)$ is the transmission function in the object plane, $o\left(\vec{r}_O\right)$ is the object distribution, $\vec{r}_O=\left(x_O,y_O,z_O\right)$ and $\vec{r}_S=\left(x_S,y_S,z_S\right)$ are the coordinates in the object respectively the screen plane, $\sigma_S$ denotes the screen plane, $z_O$ is the distance between source and object and $z_S$ the distance between source and screen. The complex-valued field in the screen plane is given by the solution of the forward propagation integral [9, 11]

$$U\left(\vec{r}_S\right)=-\frac{i}{\lambda}\iint t\left(\vec{r}_O\right)\frac{\exp\left(ikr_O\right)}{r_O}\frac{\exp(ik\left|\vec{r}_S-\vec{r}_O\right|)}{\left|\vec{r}_S-\vec{r}_O\right|}\mathrm{d}\sigma_O, \qquad (3)$$

where $\exp\left(ikr_O\right)/r_O$ describes the incident spherical wave and $\sigma_O$ denotes the object plane.

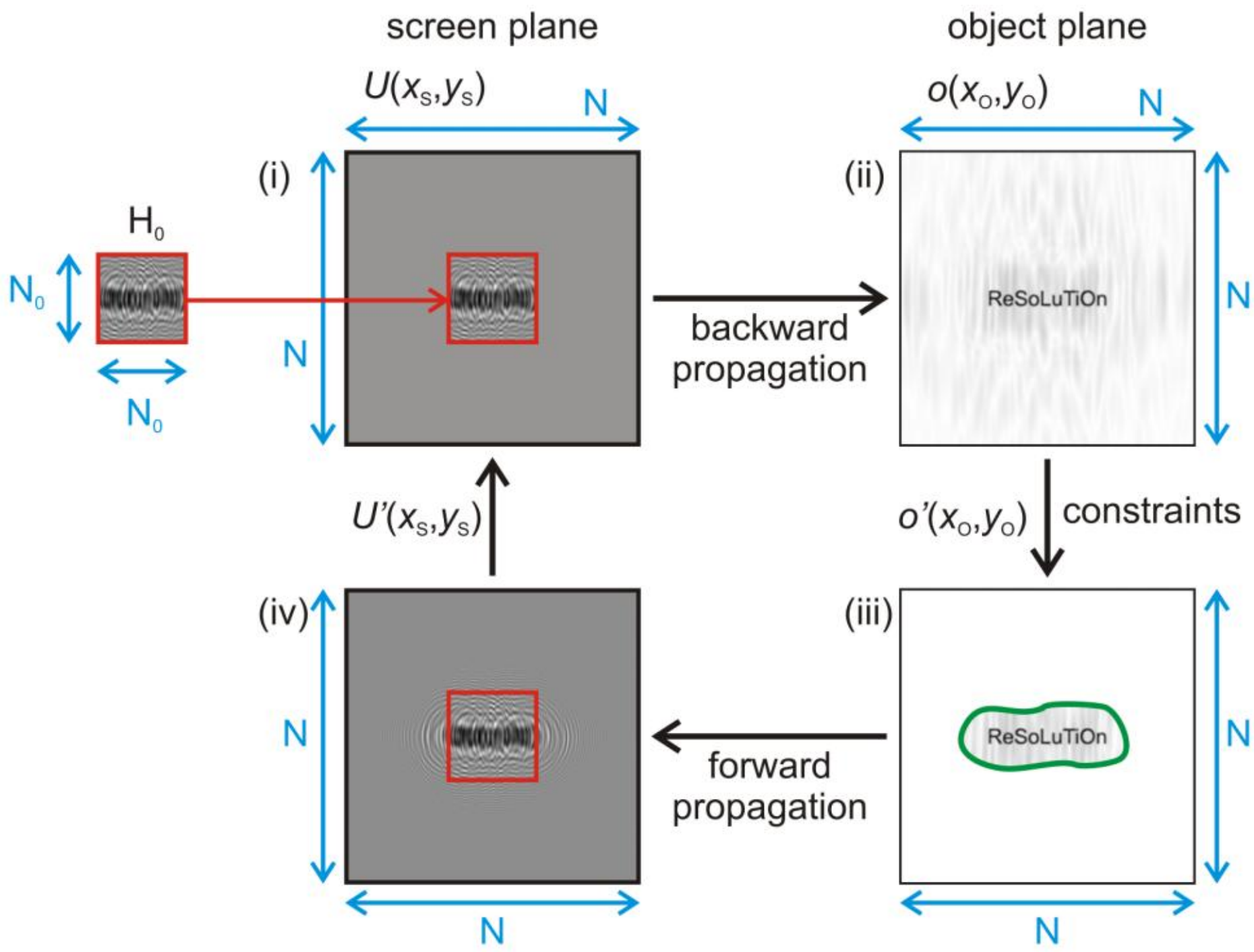

Fig. 1. Illustration describing the iterative self-extrapolation of a hologram of an object in the form of the word "ReSoLuTiOn". The iterative loop includes the steps (i)-(iv) as described in the main text. The reconstructed object amplitude distribution in (ii) and (iii) are shown in inverted intensity scale.

The iterative procedure includes the following steps as also illustrated in the flowchart diagram in Fig. 1:

(i) Formation of the input complex-valued field in the hologram plane $U(\vec{r}_S)$. The amplitude of the central $N_0 \times N_0$ part is always given by the square root of $H_0$, and the amplitude of the remaining pixels outside $H_0$ of the N×N array is updated after each iteration (for the first iteration these values are set to a constant value or are randomly distributed). The phase distribution is initially set to $kr_s$ - the phase of the known reference wave, and it is also updated after each iteration.

(ii) Back propagation to the object plane is performed by solving the propagation integral Eq. (2).

(iii) In the object plane, the following constraint is applied to the reconstructed complex-valued object distribution $o(\vec{r}_O)$. Since the object exhibits a finite size, the distribution $o(\vec{r}_O)$ is multiplied with a loose mask which sets the values outside a certain region to zero [10], see Fig.1. A second constraint is that of positive absorption based on the physical notion that the amplitude of the wave must not increase in the scattering process is also applied [9]; consequently, the pixel values where absorption is negative are set to zero. As a result, the transmission function in the object plane is updated to $t'(\vec{r}_O) = 1 + o'(\vec{r}_O)$.

(iv) By calculating the forward propagation integral given by Eq. (3), the updated complex-valued wavefront in the screen plane $U'(\vec{r}_S)$ is obtained, and its amplitude and phase distributions are the input values for the next iteration starting at step (i).

## 3. Simulated example

To demonstrate our method we first use the simulated hologram of two point scatterers, as illustrated in Fig. 2. The object area is 2 × 2 mm$^2$, sampled with 2000 × 2000 pixels, and the scatterers are separated by six pixels, Fig. 2(a). The hologram is simulated for 500 nm laser light in the inline holographic setup [9, 11], where the distance between the point source and the sample amounts to 4 mm and the distance between the point source and the hologram is 1 m; a screen size of 0.5 × 0.5 m$^2$ is used in the simulation. The resultant simulated hologram is shown in Fig. 2(b). According to Eq. 1, the resolution intrinsic to this simulated hologram amounts to 1 μm and leads to a perfectly resolved reconstruction of the two points separated by 6 μm, as shown in Fig. 2(c).

To mimic a hologram limited by detector size, the central 500 × 500 pixels region of the simulated hologram is cut out as displayed in Fig. 2(d). The resolution intrinsic to this truncated hologram amounts to 4 μm, as derived from Eq. 1. Accordingly, the reconstruction of this truncated hologram results in two smeared-out regions rather than separated points that are barely resolved, as evident from Fig. 2(e).

Next, the truncated hologram is iteratively reconstructed in the following manner. First, the surroundings of the hologram are filled with a constant background, leading to a padded record of 1000 × 1000 pixels, as shown in Fig. 2(g). For the first iteration, the phase distribution in the hologram plane is given by the phase distribution of the reference wave. During the iterative reconstruction the following constraints are applied. First, in the hologram plane, the amplitude of the central 500 × 500 pixels region is replaced by the already known data and the values of the outer pixels are updated after each iteration. Second, in the object plane, the object is confined within a limited region by setting the values outside an elliptical mask (with 23 pixels major and 13 pixels minor axis) to zero; the signal outside the mask is set to zero and then a positive absorption filter as described above is applied [9]. After 300 iterations, not only does the holographic image extrapolate itself beyond the 500 × 500 region, as evident from Fig. 2(h), but also the reconstruction of such a self-extrapolated hologram reveals that the two points are now clearly resolved, as can be seen in Fig. 2(i). A theoretical resolution of 2 μm estimated by Eq. (1) is observed in the reconstruction.

To verify that the effect of the resolution enhancement is not due to the iterative retrieval itself, we also applied the same iterative reconstruction to the non-padded 500 × 500 pixels truncated hologram. Its reconstruction after 300 iterations, shown in Fig. 2(f), does not differ much from the reconstruction shown in Fig. 2(e). While the twin image is suppressed, the two points remain barely resolved.

The reason is that when there is a larger area available in the hologram plane, the series of wavelets constituting the wavefront in the hologram plane can be fitted better. This hypothesis is confirmed by a closer look at the intensity distributions in the hologram plane. During the iterative procedure the intensity distribution in the hologram is replaced by an "experimental" distribution after each iteration and thus iteratively approaches the "experimental" distribution. The mismatch between the "experimental" amplitude distribution and the amplitude distribution updated after each iteration is quantitatively described by the error function. Figure 3 shows the intensity profiles in the hologram plane: even after 300 iterations the 500 × 500 pixels truncated hologram does not match the intensity distribution of the original hologram. But the intensity profile of the self-extrapolated 1000 × 1000 pixels hologram obtained after 300 iterations matches the original distribution perfectly. Thus, when a larger area in the hologram plane where the intensities can be varied is available, a better agreement between the "experimental" and the fitted hologram can be achieved.

Although the numerical experiments clearly show that the resolution can be enhanced by self-extrapolation of the hologram, the enhancement cannot be extended to infinity. We also tried padding from 500 × 500 pixels to the original value of 2000 × 2000 pixels, but this only marginally improves the resolution when compared to the resolution obtained by padding only up to 1000 × 1000 pixels.

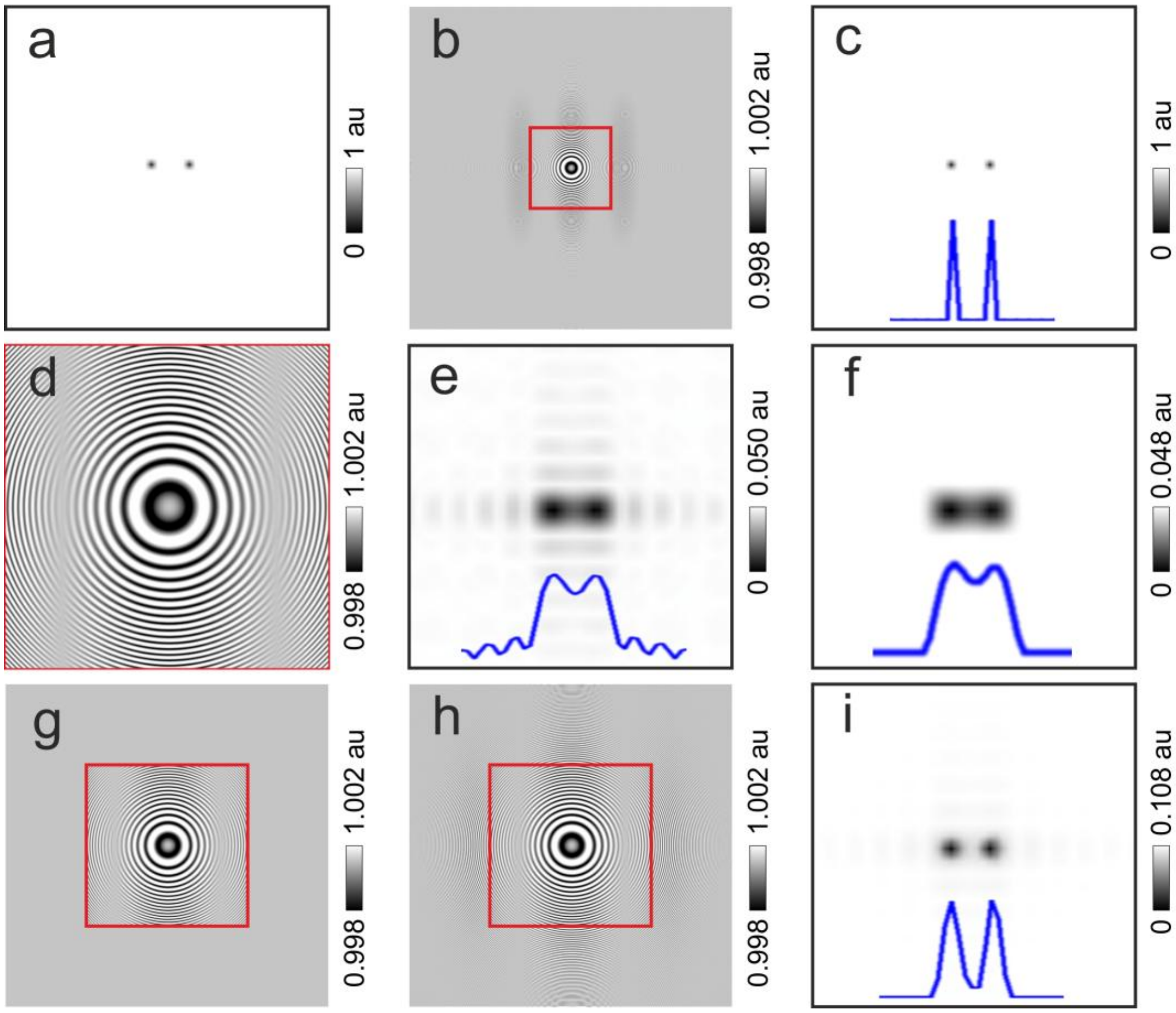

Fig. 2. Simulated example. (a) Amplitude of the transmission function of the synthetic object, consisting of two point scatterers, separated by six pixels; the central 50 × 50 pixels region is shown. (b) Simulated hologram, 2000 × 2000 pixels in size. (c) Hologram reconstruction, showing the central 50 × 50 pixels region. (d) Selected 500 × 500 pixels central region of the original hologram, as marked in (b) by a red square, and (e) its reconstruction. (f) Result of iterative reconstruction of (d) after 300 iterations. (g) The selected 500 × 500 pixels region is padded to 1000 × 1000 pixels with a constant background. (h) Self-extrapolated hologram after 300 iterations. (i) Reconstruction of the self-extrapolated hologram; the central 50 × 50 pixels region is shown. The blue curves in (c), (e), (f), and (i) show the intensity profiles of the reconstructions.

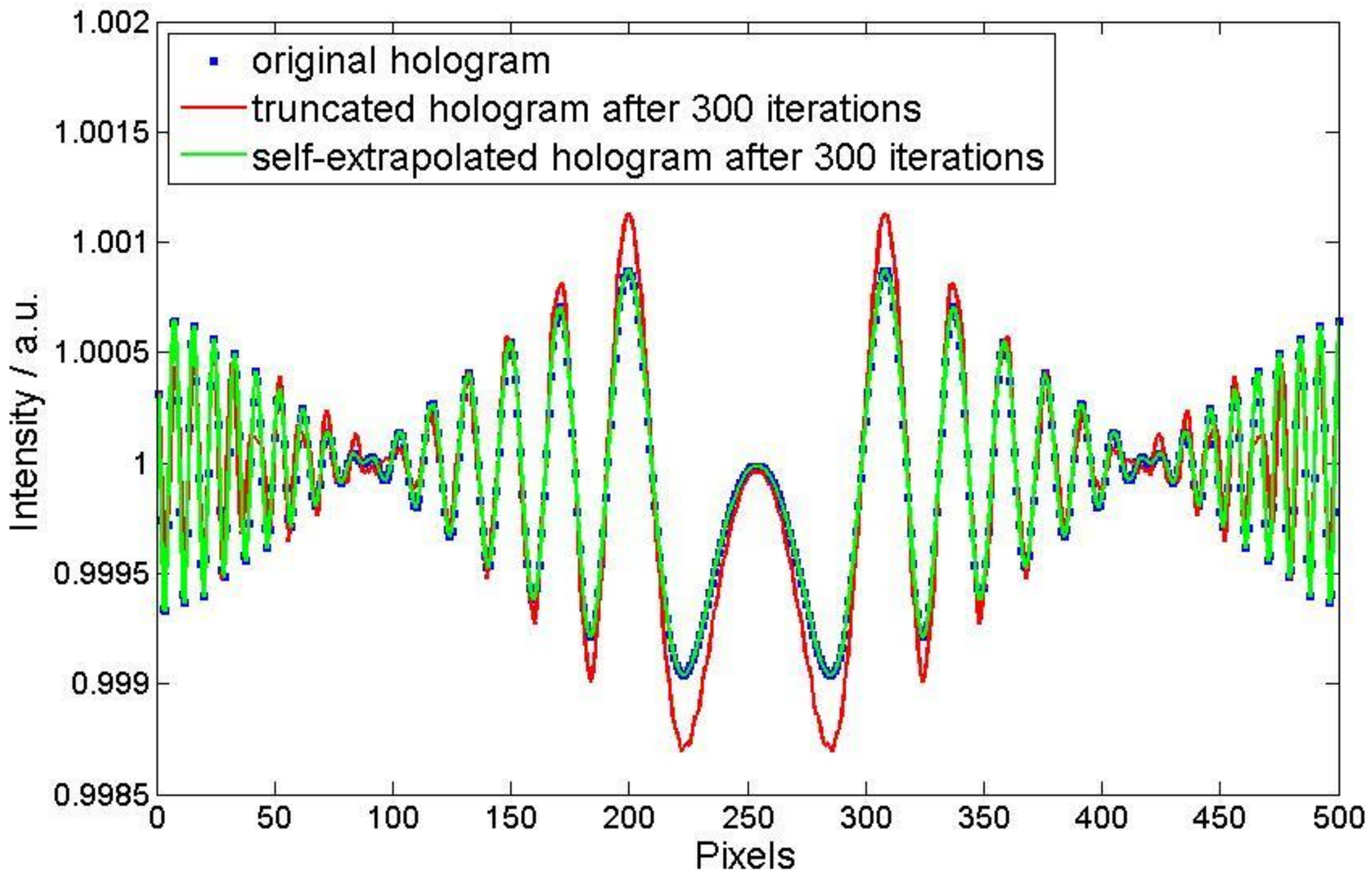

Fig. 3. Profiles of the intensity in the central region of the holograms: original hologram (blue dots), 500 × 500 pixels truncated hologram after 300 iterations (red line), and 1000 × 1000 self-extrapolated hologram after 300 iterations (green line).

**4. Experimental example**

We also performed experimental tests of our method using an optical hologram recorded in an inline scheme shown in Fig. 4. The optical hologram was recorded using 523 nm laser light; the sample was placed 0.25 mm in front of the divergent source, the distance between source and screen was 75 mm, the size of the hologram was 35 × 35 $mm^2$, and it was sampled with 1000 × 1000 pixels. The sample consisting of four circles with diameters of 5 μm each was created by focused ion beam milling in a silicon nitride membrane covered with a 200 nm platinum layer and is shown in Fig. 5(a). The normalized experimental hologram is shown in Fig. 5(b) and its reconstruction is displayed in Fig. 5(c). The normalized experimental hologram $H_0$ is obtained as following: the hologram acquired with the sample is divided by the hologram acquired without the sample [9].

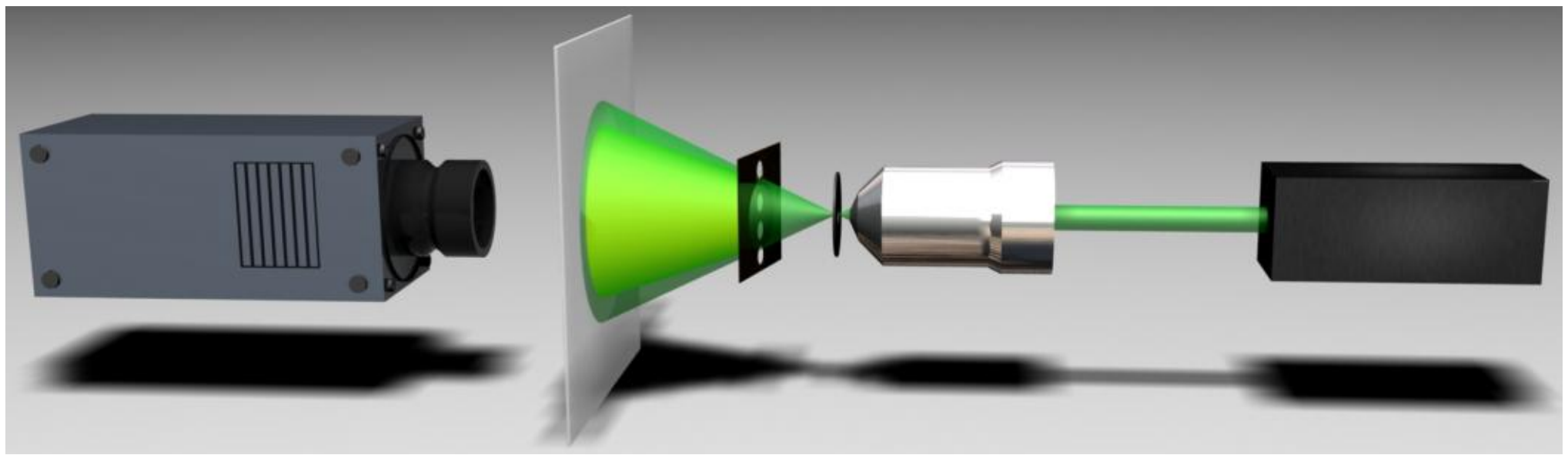

Fig. 4. Experimental scheme for recording optical inline holograms. The screen is made up of a translucent tracing paper.

When the central region of the hologram corresponding to 500 × 500 pixels is cut out (Fig. 5(d)), the reconstruction of this truncated hologram results in a blurred image of the four circles, shown Fig. 5(e). Next, the truncated hologram is padded up to 1000 × 1000 pixels, as shown in Fig. 5(f), and iteratively reconstructed as described above. After every five iterations the reconstructed amplitude was convolved (calculated using standard IMAQ Convolute.vi routine in LabView) with a 3 × 3 pixels kernel in the form of a Gaussian distribution

$$K = \begin{pmatrix} 1 & 1 & 1 \\ 1 & 4 & 1 \\ 1 & 1 & 1 \end{pmatrix} \qquad (4)$$

to smooth and suppress the accumulation of noisy peaks. The smoothing procedure was applied only after every fifth iteration to avoid excessive blurring of the reconstruction. After 100 iterations, higher order fringes emerge by self-extrapolation (see Fig. 5(g)), and the reconstruction of the self-extrapolated hologram exhibits better resolved circles, shown in Fig. 5(h).

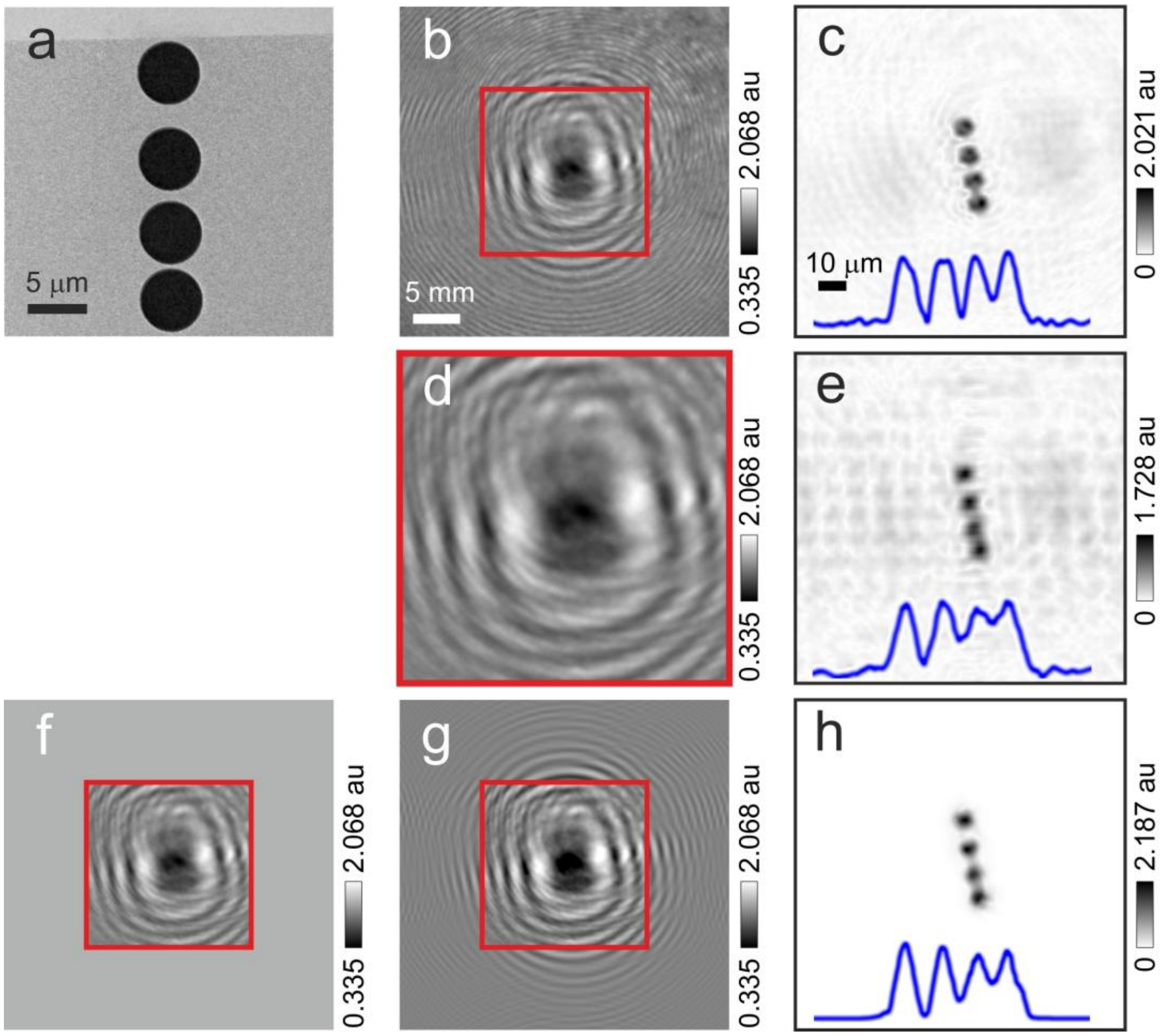

Fig. 5. Experimental verification of the method. (a) Scanning electron microscope image of the sample. (b) 1000 × 1000 pixels normalized experimental optical hologram of the sample and (c) its reconstruction; the 500 × 500 pixels central part is shown. (d) Selected 500 × 500 pixels central region of the experimental hologram, and (e) its reconstruction. (f) The 500 × 500 pixels hologram padded to 1000 × 1000 pixels. (g) The 1000 × 1000 pixels self-extrapolated hologram after 100 iterations. (h) Reconstruction of the self-extrapolated hologram; the 500 × 500 pixels central part is shown. The blue curves in (c), (e) and (h) show the intensity profiles of the reconstructions.

It is also possible to self-extrapolate a hologram when a significant low-resolution part is missing. To demonstrate it, we cropped the 500 × 500 pixels lower part of the experimental hologram and padded it up to 1000 × 1000 pixels (see Fig. 6(a)). The corresponding reconstruction presented in Fig. 6(b) does not show all four circles. Next, we applied the iterative reconstruction routine; after 300 iterations the hologram had extrapolated itself (see Fig. 6(c)), and the reconstruction clearly shows four circles (see Fig. 6(d)). Aside from the intensity variations within the reconstructed circles the agreement with the original object is good. The variations in intensity indicate the limits of the method; thus it is better to use the experimentally recorded holographic information when available.

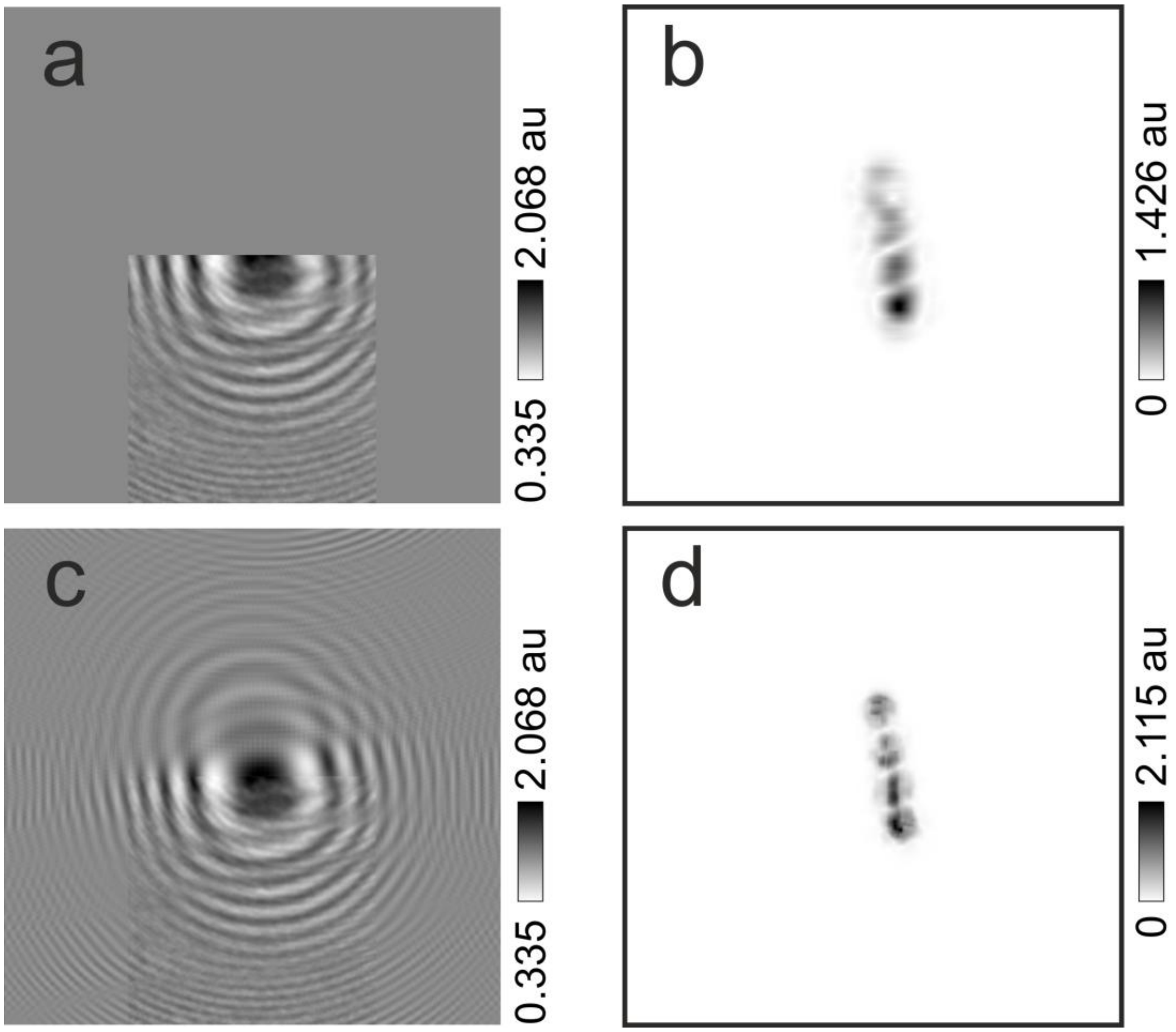

Fig. 6. Self-extrapolation of a piece of the hologram. (a) The selected 500 × 500 pixels part of the hologram is padded up to 1000 × 1000 pixels; and (b) its reconstruction; the 500 × 500 pixels central part is shown. (c) 1000 × 1000 pixels self-extrapolated hologram after 300 iterations and (d) its reconstruction; the 500 × 500 pixels central part is shown.

## 5. Conclusions

The method proposed here improves the resolution of holographic reconstructions by self-extrapolation of the digital hologram beyond the actual detector size. The method can be applied provided the following conditions are fulfilled: 1. The interference pattern is only limited by the size of the detector; in other words, there is an interference pattern at the edges of the hologram. 2. The object under study occupies a limited area and thus can be masked in the iterative reconstruction. 3. The object has a finite thickness in the z-dimension, which allows an iterative retrieval by field propagation between two planes: the hologram and object planes. 4. The available part of the hologram $H_0$ exhibits high dynamics in the intensity distribution. The latter requirement can be explained in terms of the resolution provided by the smallest scatterers whose scattered wave signal is present over the entire hologram area and needs to be detected in the available piece of the hologram. This high-resolution information eventually allows hologram self-extrapolation and retrieval of the entire object at enhanced resolution.

## Acknowledgments

We would like to thank Elvira Steinwand for the sample preparation. The Swiss National Science Foundation is gratefully acknowledged for its financial support.

```
%%%%%%%%%%%%%%%%%%%%%%%%%%%%%%%%%%%%%%%%%%%%%%
% MATLAB CODE FOR EXTRAPOLATION OF INLINE HOLOGRAM
%%%%%%%%%%%%%%%%%%%%%%%%%%%%%%%%%%%%%%%%%%%%%%
% Citation for this code/algorithm or any of its parts:
% Tatiana Latychevskaia and Hans-Werner Fink
% "Resolution enhancement in digital holography by self-extrapolation of holograms",
% Optics Express 21(6), 7726 - 7733 (2013)
%%%%%%%%%%%%%%%%%%%%%%%%%%%%%%%%%%%%%%%%%%%%%%
% The code is written by Tatiana Latychevskaia, 2013
% The version of Matlab for this code is R2010b
% The main program is a_extrapolation.m
%%%%%%%%%%%%%%%%%%%%%%%%%%%%%%%%%%%%%%%%%%%%%%
```

```
a_extrapolation.m
%%%%%%%%%%%%%%%%%%%%%%%%%%%%%%%%%%%%%%%%%%%%%%%%%%%%%%%%%%%%%%%%%%%%%%%%%%
% EXTRAPOLATION OF AN INLINE HOLOGRAM
%%%%%%%%%%%%%%%%%%%%%%%%%%%%%%%%%%%%%%%%%%%%%%%%%%%%%%%%%%%%%%%%%%%%%%%%%%
% Citation for this code/algorithm or any of its parts:
% Tatiana Latychevskaia and Hans-Werner Fink
% "Resolution enhancement in digital holography by self-extrapolation of
holograms",
% Optics Express 21(6), 7726 - 7733 (2013)
%%%%%%%%%%%%%%%%%%%%%%%%%%%%%%%%%%%%%%%%%%%%%%%%%%%%%%%%%%%%%%%%%%%%%%%%%%
% The code is written by Tatiana Latychevskaia, 2013
% The version of Matlab for this code is R2010b

close all
clear all
%%%%%%%%%%%%%%%%%%%%%%%%%%%%%%%%%%%%%%%%%%%%%%%%%%%%%%%%%%%%%
N0 = 500;      % initial number of pixels
N = 1000;      % final number of pixels
Loops = 100;   % number of iterative loops

wavelength = 532*10^(-9);   % wavelength in meter
screen = 0.035;             % screen size in meter
z = 0.075;                  % source-to-screen distance in meter
z0 = 0.252*10^(-3);         % source-to-sample distance in meter
%%%%%%%%%%%%%%%%%%%%%%%%%%%%%%%%%%%%%%%%%%%%%%%%%%%%%%%%%%%%%
% reading cropped normalized hologram
% normalized hologram = (hologram with object)/(hologram without object)

    fid = fopen('00_norm_div_cropped.bin', 'r');
    hologram = fread(fid, [N0, N0], 'real*4');
    fclose(fid);

    measured = sqrt(hologram);
%%%%%%%%%%%%%%%%%%%%%%%%%%%%%%%%%%%%%%%%%%%%%%%%%%%%%%%%%%%%%
% reading support in the object domain

    fid = fopen('00_support_object.bin', 'r');
    support = fread(fid, [N, N], 'real*4');
    fclose(fid);
%%%%%%%%%%%%%%%%%%%%%%%%%%%%%%%%%%%%%%%%%%%%%%%%%%%%%%%%%%%%%
% creating initial complex-valued field distribution in the detector plane

amplitude = ones(N,N);
phase = zeros(N,N);

spherical = spherical_wave(N, wavelength, screen, z);
% phase = angle(spherical);
field_detector = amplitude.*exp(i*phase);
%%%%%%%%%%%%%%%%%%%%%%%%%%%%%%%%%%%%%%%%%%%%%%%%%%%%%%%%%%%%%
% creating spherical wavefront term

plane0 = N*wavelength*z/screen;
spherical0 = spherical_wave(N, wavelength, plane0, z0);
%%%%%%%%%%%%%%%%%%%%%%%%%%%%%%%%%%%%%%%%%%%%%%%%%%%%%%%%%%%%%
% iterative reconstruction
```

```
    N1 = (N - N0)/2;
    N2 = (N + N0)/2;

for kk = 1:Loops

fprintf('Iteration: %d\n', kk)

% replacing the updated amplitudes with the known measured amplitudes
for ii = N1+1:N2-1
for jj = N1+1:N2-1
  amplitude(ii,jj) = measured(ii-N1,jj-N1);
end
end

field_detector = amplitude.*exp(i*phase);

% reconstruction of transmission function t
t = IFT2Dc((FT2Dc(field_detector)).*spherical0);

% filtering object function
object = t - 1;
object = object.*support;

object_amplitude = abs(object);
object_phase = angle(object);
% smoothing every 5 iterations
R = rem(kk,5);
if (R == 0)
    object_amplitude = smooth2D(object_amplitude);
end;

object = object_amplitude.*exp(i*object_phase);

t = object + 1;

% calculating complex-valued wavefront in the detector plane
field_detector_updated = FT2Dc((IFT2Dc(t)).*conj(spherical0));
amplitude = abs(field_detector_updated);
phase = angle(field_detector_updated);

end
%%%%%%%%%%%%%%%%%%%%%%%%%%%%%%%%%%%%%%%%%%%%%%%%%%%%%%%%%%%%%
% showing object
    figure
    imshow(rot90(abs(object)), [],'colormap', 1-gray);
%%%%%%%%%%%%%%%%%%%%%%%%%%%%%%%%%%%%%%%%%%%%%%%%%%%%%%%%%%%%%
% showing extrapolated hologram
for ii = N1+1:N2-1
for jj = N1+1:N2-1
  amplitude(ii,jj) = measured(ii-N1,jj-N1);
end
end
    hologram_extrapolated = amplitude.^2;
```

```
    figure
    imshow(rot90(hologram_extrapolated), [],'colormap', gray);
%%%%%%%%%%%%%%%%%%%%%%%%%%%%%%%%%%%%%%%%%%%%%%%%%%%%%%%%%%%%%%%
% saving extrapolated hologram as jpg image
        h = hologram_extrapolated;
        h = (h - min(min(h)))/(max(max(h)) - min(min(h)));
        imwrite (rot90(h), 'hologram_extrapolated.jpg');
%%%%%%%%%%%%%%%%%%%%%%%%%%%%%%%%%%%%%%%%%%%%%%%%%%%%%%%%%%%%%%%
```

spherical_wave.m

```
%%%%%%%%%%%%%%%%%%%%%%%%%%%%%%%%%%%%%%%%%%%%%%%%%%%%%%%%%%%%%%%%%%%%%%%%%%%
% COMPLEX-VALUED SPHERICAL WAVEFRONT
%%%%%%%%%%%%%%%%%%%%%%%%%%%%%%%%%%%%%%%%%%%%%%%%%%%%%%%%%%%%%%%%%%%%%%%%%%%
% Citation for this code/algorithm or any of its parts:
% Tatiana Latychevskaia and Hans-Werner Fink
% "Resolution enhancement in digital holography by self-extrapolation of
holograms",
% Optics Express 21(6), 7726 - 7733 (2013)
%%%%%%%%%%%%%%%%%%%%%%%%%%%%%%%%%%%%%%%%%%%%%%%%%%%%%%%%%%%%%%%%%%%%%%%%%%%
% The code is written by Tatiana Latychevskaia, 2013
% The version of Matlab for this code is R2010b

function [p] = spherical_wave(N, wavelength, area, z)

p = zeros(N,N);
delta = area/N;

for ii = 1:N;
    for jj = 1:N
        x = delta*(ii - N/2 -1);
        y = delta*(jj - N/2 -1);
        p(ii,jj) = exp(i*pi*(x^2 + y^2)/(wavelength*z));
    end
end;
%%%%%%%%%%%%%%%%%%%%%%%%%%%%%%%%%%%%%%%%%%%%%%%%%%%%%%%%%%%%%%%%%%%%%%%%%%%
```

smooth2D.m

```
%%%%%%%%%%%%%%%%%%%%%%%%%%%%%%%%%%%%%%%%%%%%%%%%%%%%%%%%%%%%%%%%%%%%%%%%%%
% smoothing 2d images
%%%%%%%%%%%%%%%%%%%%%%%%%%%%%%%%%%%%%%%%%%%%%%%%%%%%%%%%%%%%%%%%%%%%%%%%%%
%%%%%%%%%%%%%%%%%%%%%%%%%%%%%%%%%%%%%%%%%%%%%%%%%%%%%%%%%%%%%%%%%%%%%%%%%%
% The code is written by Tatiana Latychevskaia, 2013

function [out] = smooth2D(in)

[N N] = size(in);

      f = zeros(N,N);
      f(N/2+1,N/2+1) = 4;
      f(N/2-1+1,N/2+1) = 1;
      f(N/2+1+1,N/2+1) = 1;
      f(N/2-1+1,N/2-1+1) = 1;
      f(N/2+1,N/2-1+1) = 1;
      f(N/2+1+1,N/2-1+1) = 1;
      f(N/2-1+1,N/2+1+1) = 1;
      f(N/2+1,N/2+1+1) = 1;
      f(N/2+1+1,N/2+1+1) = 1;

out = (1/12)*abs(IFT2Dc(FT2Dc(in).*FT2Dc(f)));
%%%%%%%%%%%%%%%%%%%%%%%%%%%%%%%%%%%%%%%%%%%%%%%%%%%%%%%%%%%%%%%%%%%%%%%%%%
```

FT2Dc.m

```
%%%%%%%%%%%%%%%%%%%%%%%%%%%%%%%%%%%%%%%%%%%%%%%%%%%%%%%%%%%%%%%%%%%%%%%%%%%%
% 2d centered Fourier transform
%%%%%%%%%%%%%%%%%%%%%%%%%%%%%%%%%%%%%%%%%%%%%%%%%%%%%%%%%%%%%%%%%%%%%%%%%%%%
% Citation for this code and algorithm:
% Tatiana Latychevskaia and Hans-Werner Fink
% "Practical algorithms for simulation and reconstruction of digital in-line
holograms",
% Appl. Optics 54, 2424 - 2434 (2015)
%%%%%%%%%%%%%%%%%%%%%%%%%%%%%%%%%%%%%%%%%%%%%%%%%%%%%%%%%%%%%%%%%%%%%%%%%%%%
% The code is written by Tatiana Latychevskaia, 2002
% The version of Matlab for this code is R2010b

function [out] = FT2Dc(in)

[Nx Ny] = size(in);

f1 = zeros(Nx,Ny);

for ii = 1:Nx
    for jj = 1:Ny
         f1(ii,jj) = exp(i*pi*(ii + jj));
    end
end

FT = fft2(f1.*in);

out = f1.*FT;
%%%%%%%%%%%%%%%%%%%%%%%%%%%%%%%%%%%%%%%%%%%%%%%%%%%%%%%%%%%%%%%%%%%%%%%%%%%%
```

IFT2Dc.m

```
%%%%%%%%%%%%%%%%%%%%%%%%%%%%%%%%%%%%%%%%%%%%%%%%%%%%%%%%%%%%%%%%%%%%%%%%%%%%%%
% 2d centered inverse Fourier transform
%%%%%%%%%%%%%%%%%%%%%%%%%%%%%%%%%%%%%%%%%%%%%%%%%%%%%%%%%%%%%%%%%%%%%%%%%%%%%%
% Citation for this code and algorithm:
% Tatiana Latychevskaia and Hans-Werner Fink
% "Practical algorithms for simulation and reconstruction of digital in-line
holograms",
% Appl. Optics 54, 2424 - 2434 (2015)
%%%%%%%%%%%%%%%%%%%%%%%%%%%%%%%%%%%%%%%%%%%%%%%%%%%%%%%%%%%%%%%%%%%%%%%%%%%%%%
% The code is written by Tatiana Latychevskaia, 2002
% The version of Matlab for this code is R2010b

function [out] = IFT2Dc(in)

[Nx Ny] = size(in);

f1 = zeros(Nx,Ny);

for ii = 1:Nx
    for jj = 1:Ny
        f1(ii, jj) = exp(-i*pi*(ii + jj));
    end
end

FT = ifft2(f1.*in);

out = f1.*FT;
%%%%%%%%%%%%%%%%%%%%%%%%%%%%%%%%%%%%%%%%%%%%%%%%%%%%%%%%%%%%%%%%%%%%%%%%%%%%%%
```